\documentclass[8.5pt]{revtex4}

\usepackage{graphicx}

\begin{document}

\title{Structure of graphene oxide: thermodynamics
versus kinetics}

\author{Ning Lu}
\author{Zhenyu Li}
\thanks{Corresponding author. E-mail: zyli@ustc.edu.cn}
\author{Jinlong Yang}

\affiliation{Hefei National Laboratory for Physical Sciences at
     Microscale,  University of Science and Technology of
     China, Hefei,  Anhui 230026, China}

\maketitle

\noindent \normalsize{Graphene oxide (GO) is an important
intermediate to prepare graphene and it is also a versatile material
with various applications. However, despite its importance, the detailed
structure of GO is still unclear. For example, previous
theoretical studies based on energetics have suggested that hydroxyl
chain is an important structural motif of GO, which, however, is
found to be contrary to nuclear magnetic resonance (NMR)
experiment. In this study, we check both
thermodynamic and kinetic aspects missed previously. First principles thermodynamics
gives a free energy based stability ordering similar to that based on energetics, and hydroxyl chain is thus
thermodynamically still favorable. At the same time, by checking the calculated
vibrational frequencies, we note that hydroxyl chain structure is
also inconsistent with infrared experiment. Therefore, kinetics during GO synthesis
is expected to make an important role in GO structure. Transition state calculations
predict large energy barriers between local minima, which
suggests that experimentally obtained GO has a kinetically constrained structure. } \vspace{0.5cm}


Recently, an intense research interest has been attracted by graphene
oxide (GO).\cite{Dreyer201028} Reduction of GO is a promising way to massively product
graphene sample. \cite{Stankovich200682,
Gilje200794, Li200801, Gao200903}  At the same time, many
applications of GO have been demonstrated, such as electronic device, \cite{Wu200801, Wang201009, Kim201003}
chemical catalyst, \cite{Scheuermann200962, Eda201005} hydrogen storage, \cite{Wang200920, Psofogiannakis200933} and
functional materials. \cite{Dikin200757, Burress201002} Both to
improve the sample quality of graphene from GO reduction and to better
utilize GO as a new material, it is very desirable to understand
the atomic details of GO structure.

A strong research effort has been devoted to the GO structure study,
both theoretically \cite{Boukhvalov200897, Lahaye200935, Li200920,
Yan200902, Wang200995, Wang201006, Yan201003, Lu201002, Paci200799,
Ghaderi201025, Zhang200905} and experimentally. \cite{Gao200903,
Lerf199877, Cai200815, Casabianca201072, Saxena201033, Lee201023} Based on
nuclear magnetic resonance (NMR) experiment, it is believed that GO mainly
has hydroxyl (-OH) and epoxy (-O-) groups on the basal plane and
carboxylic acid groups (-COOH) at edge sites. \cite{ Lerf199877,
Cai200815, Casabianca201072} Hydroxyl and epoxy groups are in close proximity,
while sp$^2$ carbon prefers to form small aromatic areas. \cite{Cai200815, Lu201002}
More details about GO structure, such as possible structural motifs, are still
under study.

Theoretically, many GO structure models based on small supercells
have been constructed and asserted by comparing energies.
\cite{Boukhvalov200897, Lahaye200935, Yan201003, Wang201006, Yan200902}
Important GO structure characteristics, such as the proximity of hydroxyl and
epoxy groups, has been obtained in these studies.
However, some controversies between theory and experiment exist. For example, although
partially oxidized stable GO samples are routinely obtained in experiment,
computational energetics suggests that fully covered GO is the most
stable structure.\cite{Yan200902} Another important conclusion from theoretical
studies is that hydroxyl chain in GO is a very stable structure.
\cite{Yan200902, Wang200995, Wang201006, Yan201003} This result looks
reasonable, since hydrogen bonds formed in this structure will strongly
stabilize the system. However, our previous NMR simulation found that
hydroxyl chain gives a too large $^{13}$C chemical shift, and it
thus should not be an important structural motif of GO.
\cite{Lu201002}

A possible reason for these discrepancies is from the thermodynamics:
energetically favorable structure may not be the
thermodynamically most stable one. For example, although GO is
synthesized in solution, previous theoretical studies did not include
solvation effects. Since hydroxyl group can also form hydrogen bond
with solvent molecules, the stabilization effect of hydroxyl chain
may be weakened by taking solution into account. Also, the
phonon contribution to free energy has not been considered in
previous theoretical studies, which may also alter the stability ordering.
In this study, free energies of different GO structure models
have been compared. However, the trend
obtained from energetics is found to be qualitatively correct in the
thermodynamics point of view. Therefore, we also consider the
kinetics of GO structure evolution. We find that it is difficult
to relax the GO structure from one local minimum to another.
Therefore, GO may exist with a kinetically constrained structure, which
explains the controversies between theory and experiment well.


Free energy ($G$) is calculated with density functional theory (DFT)
implemented in the DMOL$^{3}$ package \cite{Delley199008, Delley200356}
under the generalized gradient approximation\cite{Perdew199677}
\begin{equation}
G=E_0+\Delta G_{sol}+E_{ZPE}+k_BT\sum_iln(1-e^{-\hbar\omega_i/k_BT})
\end{equation}
where $\Delta G_{sol}$ is the free energy of solvation, and its
electrostatic part is calculated using COMSO water solution model. \cite{Klamt199399, Delley200617}
$E_{ZPE}$ is the zero point energy correction, and $\omega_i$ is
the phonon frequency at the zone center, which is calculated
by finite difference using a large supercell. Minimum-energy pathway for elementary
reaction step is computed using the nudged elastic band (NEB)
method \cite{Henkelman200078} implemented in the Vienna Ab initio
Simulation Package (VASP). \cite{n8, n9}

As a benchmark, we first consider isolated epoxy group and hydroxyl
group using a 5$\times$5 graphene supercell. With a single epoxy group, the
optimized structure gives a C-O bond length of 1.46 \AA, and the distance
between the two oxidized C atoms is 1.51 \AA. The O-H bond length in
an isolated hydroxyl group is 0.98 \AA, and the corresponding C-O
bond length is 1.49 \AA. Vibrational frequency of O-H stretch is
3659.1 $cm^{-1}$ in the isolated hydroxyl group. Geometric parameters
and vibrational frequency reported here agree well with previous
study. \cite{Yan201003}  The calculated $\Delta G_{sol}$ is only -0.08 eV
for isolated epoxy group, while it is -0.25 eV for isolated hydroxyl
group, which is consistent with the stronger polarization of
hydroxyl group.

\begin{figure}[tbhp]
\centering
\includegraphics[width=8cm]{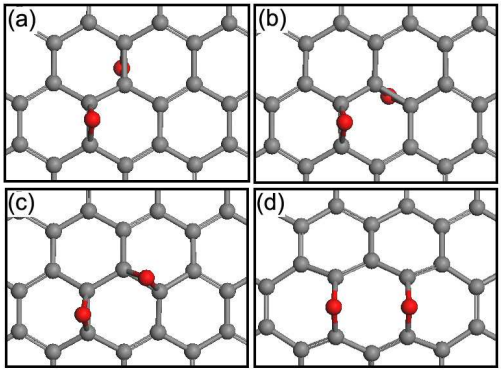}
\caption{Different configurations of two epoxy groups. A
5$\times$5 supercell is using in calculation, which is not shown in the figure. Carbon is in gray and
oxygen is in red.} \label{fig:2o}
\end{figure}

To check the trends of energies and free energies ordering for
different GO structures, we first consider two epoxy groups (2O).
Four representative configurations including the two with lowest
energies \cite{Wang201006,Yan201003} are studied (Figure
\ref{fig:2o}). Their relative energies are listed in Table
\ref{tb2:2o}(1a-1d). Structures a and b have almost the same energy,
which is 0.10 and 0.26 eV lower than that of c and d, respectively.
This can be understood from a consideration of tension caused by epoxy
groups. For structures a and b, tensions from both sides will
compensate each other to decrease the total energy. Aligned epoxy
groups prefer to break underlying C-C bonds, \cite{Li200920} as shown in
Figure \ref{fig:2o}d. The solvation free energy varies from -0.08 to
-0.18 eV. With the large unit cell used here, zero point energy
correction is more than 8 eV, but their differences are small.
Finally, the relative free energies at 300K is similar to relative
energies, with the same ordering.

\begin{table}[bth]
\caption{Energy ($E_0$), solvation free energy ($\Delta G_{sol}$),
zero point correction ($E_{ZPE}$),
and free energy at 300 K ($G(300)$) of different GO models.
$E_0$ and $G(300)$ are listed with the most stable
configuration as the reference. All values are
in eV. } \label{tb2:2o}
\begin{tabular}{ccccccccccccccccccccccccc}
 \hline\hline
       &  &$E_0$ & $\Delta G_{sol}$ & $E_{ZPE}$& $G(300)$    \\
 \hline
              2O     & 1a  & 0.00 & -0.14 & 8.72 & 0.00   \\
                     & 1b  & 0.00 & -0.16 & 8.75 & 0.02   \\
                     & 1c  & 0.10 & -0.18 & 8.69 & 0.04   \\
                     & 1d  & 0.26 & -0.08 & 8.70 & 0.35   \\
\hline
             2OH     & 2a  & 0.00 & -0.30 & 9.37 & 0.00    \\
                     & 2b  & 0.40 & -0.24 & 9.30 & 0.39    \\
                     & 2c  & 0.70 & -0.31 & 9.30 & 0.61    \\
                     & 2d  & 1.37 & -0.39 & 9.27 & 1.19    \\
\hline
         O+2OH       & 3a  & 0.00 & -0.28 & 9.45 & 0.00     \\
                     & 3b  & 0.03 & -0.26 & 9.44 & 0.03     \\
                     & 3c  & 0.68 & -0.37 & 9.45 & 0.57     \\
                     & 3d  & 0.86 & -0.46 & 9.39 & 0.61     \\
\hline
       O-full        & 4a  & 0.00 & -0.34 & 5.12 & 0.00      \\
                     & 4b  & 3.25 & -0.39 & 5.08 & 3.17      \\
\hline
       OH-full       & 4c  & 0.00 & -0.03 & 2.33 & 0.00       \\
                     & 4d  & 0.13 & -0.02 & 2.30 & 0.07       \\
\hline
       O-OH-full     & 5a  & 0.00 & -0.19 & 4.02 & 0.00       \\
                     & 5b  & 0.19 & -0.19 & 4.03 & 0.20       \\
\hline
       O-OH-sp$^2$   & 5c  & 0.00 & -0.18 & 3.83 & 0.00      \\
                     & 5d  & 0.12 & -0.24 & 3.84 & 0.03       \\

 \hline\hline
\end{tabular}
\end{table}

\begin{figure}[tbhp]
\centering
\includegraphics[width=8cm]{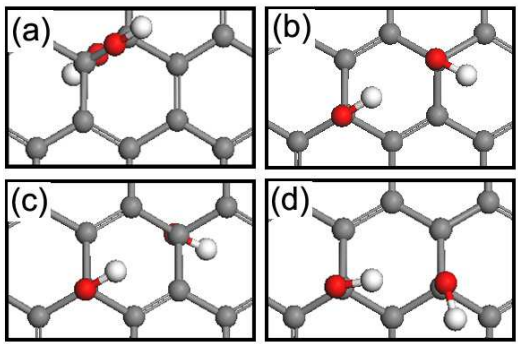}
\caption{Different configurations of the two hydroxyl groups.
A 5$\times$5 supercell is used in calculation, which is not shown in the figure. Carbon is in gray,
oxygen is in red, and hydrogen is in white.} \label{fig:2oh}
\end{figure}

We then study 2OH models (Figure \ref{fig:2oh}) with two hydroxyl groups.
Their relative energies are listed in Table \ref{tb2:2o} (2a-2d). Our calculation
shows the 1,2-hydroxyl pair has the lowest energy, which has been widely
recognized in previous studies. \cite{Yan200902, Wang200995} Free energies
follow the same order as energies, and their relative changes are small.
In structures b and d, hydrogen bonds are formed. The O-H bonds involving in
hydrogen bonds are slightly elongated, leading to two softer O-H stretching
mode with frequencies about 3450 $cm^{-1}$. For other
O-H bonds, the corresponding frequencies are between 3640 and 3720
$cm^{-1}$.

\begin{figure}[tbhp]
\centering
\includegraphics[width=8cm]{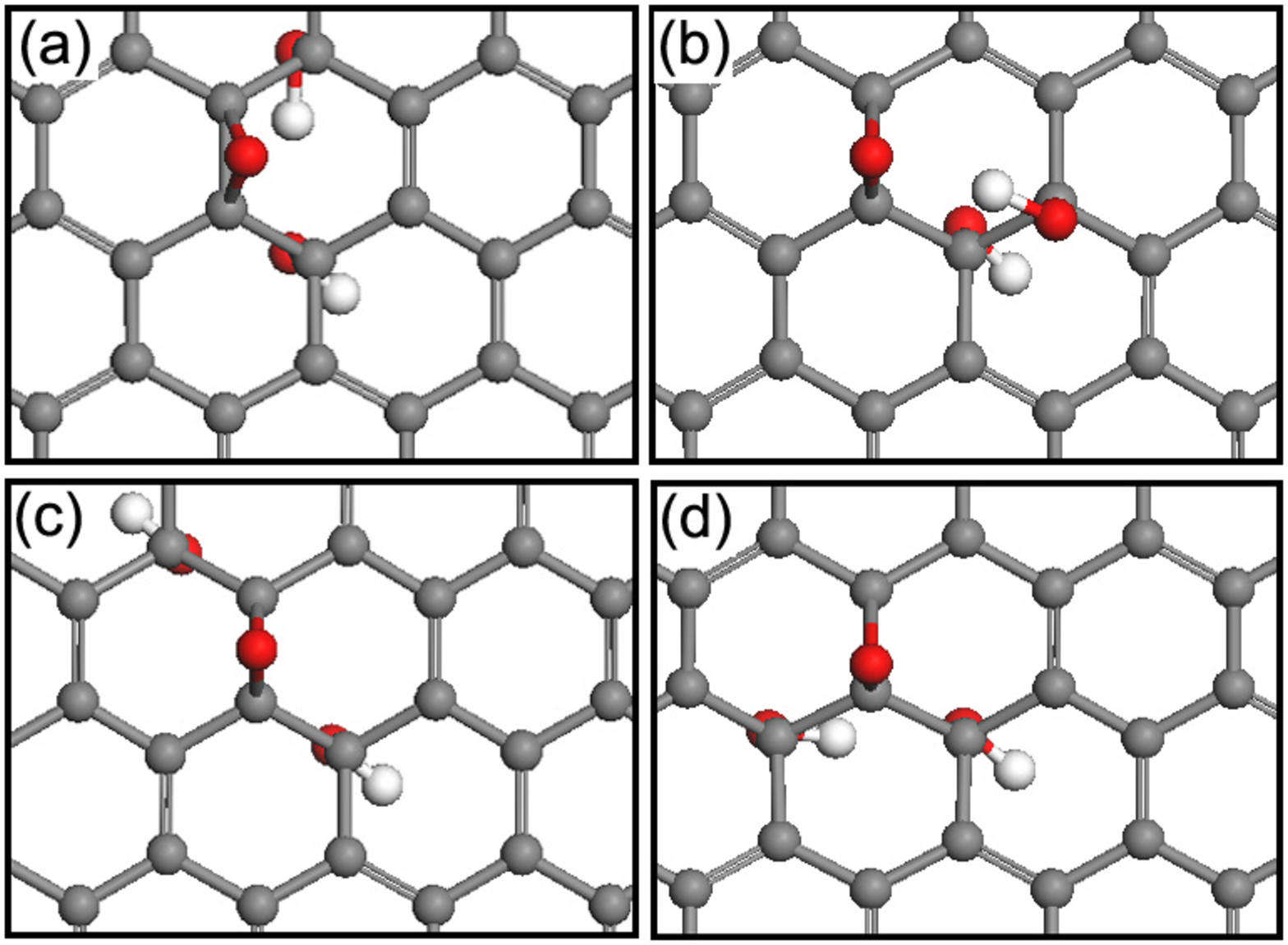}
\caption{Different configurations of one epoxy group plus two
hydroxyl groups. A 5$\times$5 supercell is used in calculation, which is not shown in the figure.
Carbon is in gray, oxygen is in red, and hydrogen is in white.}
\label{fig:o+2oh}
\end{figure}

Considering the proximity of hydroxyl and epoxy groups in real GO structure, we
discuss several typical O+2OH models (Figure \ref{fig:o+2oh}). Their relative energies
are listed in Table \ref{tb2:2o} (3a-3d). In structure a, two hydroxyl groups are attached to
carbon atoms directly adjacent to epoxide at the opposite side of
the carbon plane, which is the most stable. Structure b, with an epoxy group and a
neighboring 1,2-hydroxyl pair, has almost the same energy (0.03 eV higher).
Structures c and d have much higher energies. Calculated free energies have the
same trend. Hydrogen
bond induced softening of O-H stretching mode is also observed similar to the 2OH case.

\begin{figure}[tbhp]
\centering
\includegraphics[width=8cm]{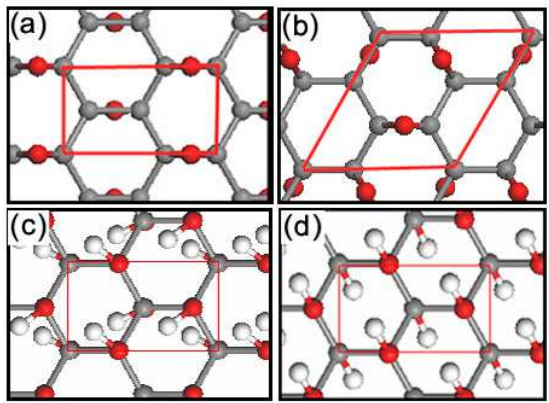}
\caption{(a) and (b) Different configurations of fully
oxidized graphene epoxide. (c) and (d) Different
configurations of fully oxidized GO with hydroxyl groups only. The unit
cell used in calculation is marked in red. Carbon is in gray, oxygen is in red, and
hydrogen is in white.} \label{fig:full-o+oh}
\end{figure}

Besides these isolated-oxidation-group models, we also consider fully oxidized
models, which goes to another limit. The fully oxidized graphene epoxide (O-full) has two
configurations, which are shown in Figure
\ref{fig:full-o+oh}a and b. With a 2$\times$3 supercell for structure a
and a 2$\times$2 supercell for structure b, we have the same number of atoms
in both structures. The calculated relative energy and relative free energy are
still very similar.

Two models with 100\% hydroxyl coverage (OH-full) are
constructed. One (Figure \ref{fig:full-o+oh}c) contains hydroxyl
chains, while the other (Figure \ref{fig:full-o+oh}d) does not. As
shown in Table \ref{tb2:2o} (4c-4d), structure c with hydroxyl chains
is 0.13 eV lower in energy. Its free energy is also 0.07 eV lower.
Strong hydrogen bonds in structure c elongate O-H bonds to about
1.0 \AA\, and decrease O-H stretching frequencies to
3050-3270 $cm^{-1}$. In contrast, O-H stretching in structure d has frequencies from 3360
to 3500 $cm^{-1}$.

\begin{figure}[tbhp]
\centering
\includegraphics[width=8cm]{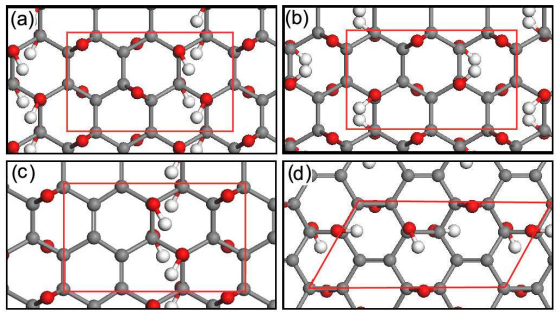}
\caption{(a)and (b) Different configurations of fully
oxidized GO with both epoxy and hydroxyl groups. (c) and
(d) Different configurations of partially oxidized GO with
both epoxy and hydroxyl groups. The unit
cell used in calculation is marked in red. Carbon is in gray, oxygen is in
red, and hydrogen is in white.} \label{fig:full+par}
\end{figure}

Structure a in Figure \ref{fig:full+par} is a stable fully-oxidized GO model (O-OH-full) with hydroxyl chains
proposed by Yan et al. \cite{Yan200902} based on energetics, which
is 0.19 eV lower in energy than structure b without hydroxyl chain.
When free energy is considered
instead, structure a is still 0.20 eV more stable than structure b.
O-H bonds in hydroxyl chain are elongated to about 1.01 \AA\ due to the formed hydrogen bonds.
Corresponding O-H stretching frequencies are generally between 2800-3100 $cm^{-1}$. These
frequencies for structure b are around 3450 $cm^{-1}$.
Since the experiment value of O-H stretching
frequency of GO is about 3400 $cm^{-1}$ \cite{Choi201007, Nethravathi200840} or higher\cite{Szabo200640}, our frequency
calculations also suggested that hydroxyl chain is not a GO
structure motif.

As another test, partially oxidized structures (O-OH-sp$^2$) with the ratio of
C(sp$^2$)/C(-O-)/C(-OH) equal to 1:1:1 are considered (Figure
\ref{fig:full+par}c and \ref{fig:full+par}d). As shown in Table \ref{tb2:2o} (5c-5d),
structure c with hydroxyl chains is 0.12 eV lower in energy than
structure d and 0.04 eV lower in free energy. Although the
difference becomes smaller when free energy is considered, it is
safe to conclude that hydroxyl chain is thermodynamically still a
very stable structure. The O-H stretching frequencies in structure c are
among 2800-3100 $cm^{-1}$, much smaller than experimental value.
Those in structure d are around 3580 $cm^{-1}$.

Since hydroxyl chain gives too large chemical shift and too small O-H stretching
frequency compared to experimental values, while it is still very stable with both solvation and
phonon contributions to free energy taken into account, it is more
likely the formation of hydroxyl chain is prohibited by kinetics
during the formation of GO. To check the possibility of experimentally
obtained GO as a metastable state, we consider the
oxidation-group diffusion in GO.

\begin{figure}[tbhp]
\centering
\includegraphics[width=8cm]{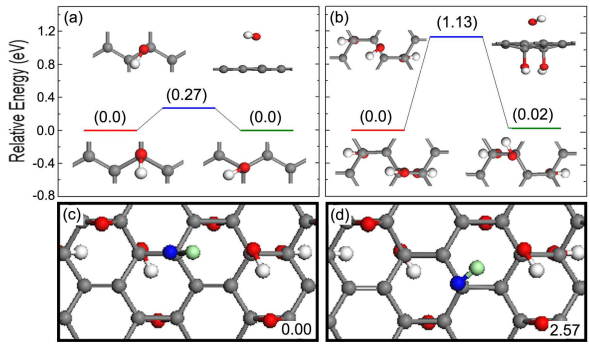}
\caption{(a) and (b) Energy profile for hydroxyl diffusion. Insets are
optimized structure of the initial, transition, and final states.
A 5$\times$5 supercell is used in calculation, which is not shown
in the figure. (c) Initial state of hydroxyl diffusion in a partially
oxidized GO, and its corresponding (d) final state, where relative
energies are marked.
Carbon is in gray, oxygen is in red, and hydrogen is in white.
In (c) and (d), atoms involving in the diffusion in marked in different colors
(blue for O and green for H).} \label{fig:ohdiff}
\end{figure}

As shown in Figure \ref{fig:ohdiff}a, isolated hydroxyl group is
easy to diffuse on pristine graphene surface \cite{Ghaderi201025} with an energy barrier
only 0.27 eV. The C-O distance is 1.51
\AA\ in the initial state and it is elongated to 2.51 \AA\ in the
transition state. When we add an additional neighboring hydroxyl
group as shown in Fig \ref{fig:ohdiff}b, the diffusion barrier
increases to 1.13 eV. Therefore, the diffusion barrier of hydroxyl
group strongly depends on its environment. It is thus important to
check the hydroxyl mobility in a realistic GO model with a certain
epoxy and hydroxyl group concentration. With the model shown in
Figure \ref{fig:ohdiff}c, the diffusion
barrier becomes as large as 2.89 eV. This is because that a stable GO
structure requires a delicate balance between positions of hydroxyl
and epoxy groups to minimize stress. Typically, moving a hydroxyl
group at the same time fixing other groups will lead to a large
change in energy. In our case, the finial state is 2.57 eV higher in
energy than the initial state, which leads to a very high diffusion
barrier.

\begin{figure}[tbhp]
\centering
\includegraphics[width=8cm]{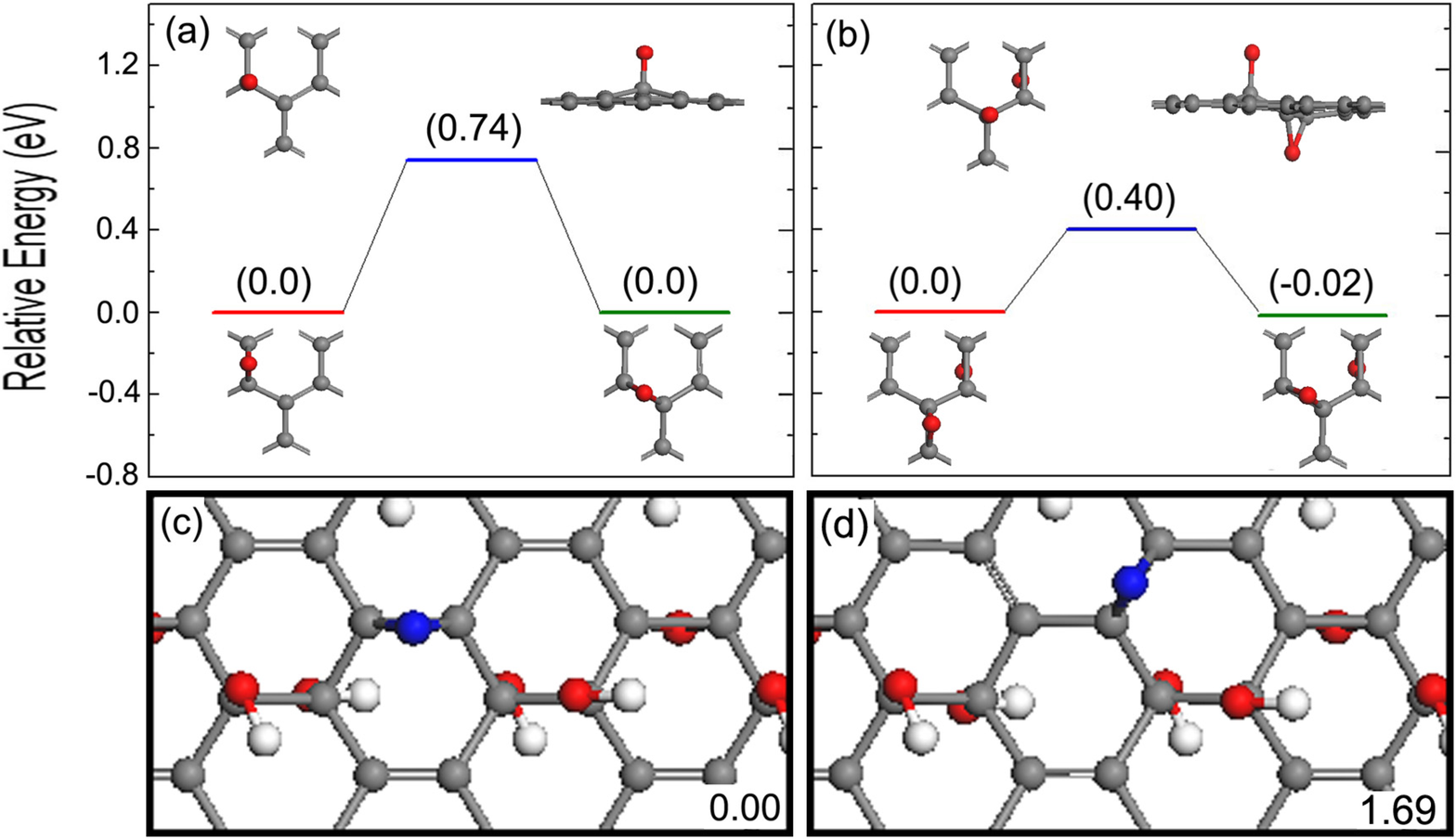}
\caption{(a) and (b) Energy profile for epoxy group diffusion. Insets are
optimized structure of the initial, transition, and final states.
A 5$\times$5 supercell is used in calculation, which is not shown
in the figure. (c) Initial state of epoxy group diffusion in a partially
oxidized GO, and its corresponding (d) final state, where relative
energies are marked.
Carbon is in gray, oxygen is in red, and hydrogen is in white.
In (c) and (d), atoms involving in the diffusion in marked in different colors
(blue for O and green for H).} \label{fig:odiff}
\end{figure}

The diffusion barrier for an isolated epoxy group is 0.74 eV,
similar to the value (0.9 eV) obtained by Li \textit{et al.}
previously with cluster models. \cite{Li200601} When a neighboring
epoxy group is added (Figure \ref{fig:odiff}b), the diffusion
barrier becomes 0.40 eV, which also demonstrates a strong environment
dependence. Similar to the hydroxyl diffusion case, in a realistic
GO model (Figure \ref{fig:odiff}c), a very
large diffusion barrier (2.0 eV) is obtained due to the energy
difference between the initial and final states.

\begin{figure}[tbhp]
\centering
\includegraphics[width=8cm]{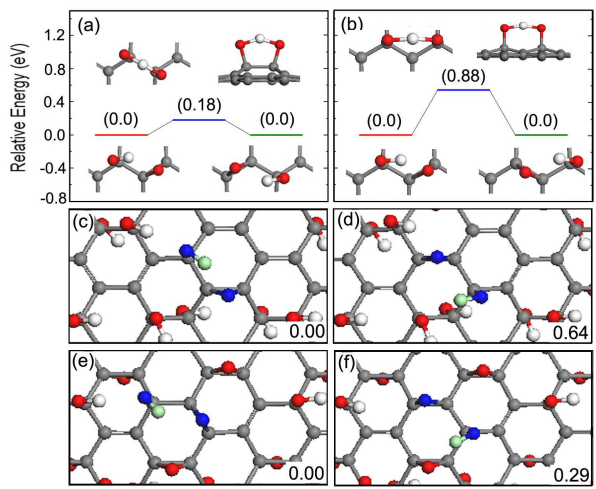}
\caption{(a) and (b) Energy profile for H diffusion from a hydroxyl group
to a neighboring epoxy group. Insets are
optimized structure of the initial, transition, and final states.
A 5$\times$5 supercell is used in calculation, which is not shown
in the figure. (c) Initial state of H diffusion with 1,2-transition state in a partially
oxidized GO, and its corresponding (d) final state, where relative
energies are marked. (e) Initial state of H diffusion with 1,3-transition state in a partially
oxidized GO, and its corresponding (f) final state, where relative
energies are marked.
Carbon is in gray, oxygen is in red, and hydrogen is in white.
In (c) and (d), atoms involving in the diffusion in marked in different colors
(blue for O and green for H).} \label{fig:hdiff}
\end{figure}

Exchange of a hydroxyl group and a neighboring epoxy group can be realized by a
H diffusion between them. There are two different kinds of transition states,
which leads to different final states. The first kind of transition states
occupy 1,2-sites (Figure \ref{fig:hdiff}a) and the second one is on 1,3-sites (Figure
\ref{fig:hdiff}b). In the first case, the
diffusion barrier is very small (0.18 eV), as also reported previously with
cluster models. \cite{Psofogiannakis200933}
In the second case, we obtain a much higher diffusion barrier of 0.88
eV. The longer distance between H and O in the transition
state leads to higher diffusion barrier. When a realistic GO model is used, the diffusion
barrier increase to 1.33 and 1.95 eV for the 1,2- and 1,3-cases,
respectively.

According to our calculations, although it is possible to design
some delicate structures, in which the barrier of oxidation group
diffusion is low, generally the diffusion barrier is very high at
least in one direction. Therefore, a global relaxation of GO
structure is very difficult. GO may easily relax to a local minimum,
as evidenced also by molecular dynamics simulations, \cite{Paci200799}
but different local minima are expected to be separated by large barriers.
Experimentally available GO thus typically has a kinetically
constrained metastable structure, which explains why spectroscopic signal of
thermodynamically stable hydroxyl chain structure is not observed in
experiment. This result can also explain why homogenous phase is
formed in partially oxidized GO, although phase separated structure
is more stable.\cite{Wang201006}

In summary, both thermodynamic and kinetic aspects of GO structure have been
considered from first principles. By taking the solvation effect and
phonon contribution into account, we calculate the free energies
of different GO structure models. We find that hydroxyl chain is really a
very stable structure. However, the calculated vibrational
frequencies and also previously obtained chemical shifts of
this structure is not observed in infrared and NMR experiments.
Therefore, kinetics during the synthesis of GO must have made an
important role in GO structure. We then study the diffusion
of oxidation groups in GO. Typically, there is a very large diffusion
barrier. Therefore, experimentally available GO will has a kinetically constrained structure.
To understand GO structure better, more studies on the oxidation
process of graphite is highly desirable.

\textbf{Acknowledgements.} The authors are grateful to Prof. Zhi-Pan Liu for helpful discussion.
This work is partially supported by NSFC (20933006 and 20803071), by MOE
(FANEDD-2007B23 and NCET-08-0521), by MOST (2011CB921404), by CAS
(KJCX2-YW-W22), and by USTC-SCC, SCCAS, and Shanghai Supercomputer Centers.

\end{document}